\documentclass[preprint2]{aastex}

\usepackage{amsmath}
\usepackage{graphicx}
\usepackage{epsfig,psfrag,epic,eepic}
\usepackage{textcomp}
\usepackage{times}
\usepackage{longtable}

\slugcomment{Published in the Astrophysical Journal}

\shorttitle{Non-detection of Fomalhaut b}
\shortauthors{Janson et al.}

\begin{document}

\title{Infrared Non-detection of Fomalhaut b -- Implications for the Planet Interpretation}

\author{Markus Janson\altaffilmark{1,5}, 
Joseph C. Carson\altaffilmark{2}, 
David Lafreni{\`e}re\altaffilmark{3}, 
David S. Spiegel\altaffilmark{4}, 
John R. Bent\altaffilmark{2}, 
Palmer Wong\altaffilmark{2}
}

\altaffiltext{1}{Princeton Univ., USA; \texttt{janson@astro.princeton.edu}}
\altaffiltext{2}{College of Charleston, Charleston, USA}
\altaffiltext{3}{University of Montreal, Montreal, Canada}
\altaffiltext{4}{Institute for Advanced Studies, Princeton, USA}
\altaffiltext{5}{Hubble fellow}

\begin{abstract}\noindent
The nearby A4-type star Fomalhaut hosts a debris belt in the form of an eccentric ring, which is thought to be caused by dynamical influence from a giant planet companion. In 2008, a detection of a point-source inside the inner edge of the ring was reported and was interpreted as a direct image of the planet, named Fomalhaut~b. The detection was made at $\sim$600--800~nm, but no corresponding signatures were found in the near-infrared range, where the bulk emission of such a planet should be expected. Here we present deep observations of Fomalhaut with \textit{Spitzer}/IRAC at 4.5~$\mu$m, using a novel PSF subtraction technique based on ADI and LOCI, in order to substantially improve the \textit{Spitzer} contrast at small separations. The results provide more than an order of magnitude improvement in the upper flux limit of Fomalhaut~b and exclude the possibility that any flux from a giant planet surface contributes to the observed flux at visible wavelengths. This renders any direct connection between the observed light source and the dynamically inferred giant planet highly unlikely. We discuss several possible interpretations of the total body of observations of the Fomalhaut system, and find that the interpretation that best matches the available data for the observed source is scattered light from transient or semi-transient dust cloud.
\end{abstract}


\section{Introduction}
\label{sec_intro}

Direct imaging of exoplanets has been a field in rapid development over the past few years, with detections of several planets \citep[e.g.][]{lagrange2010,marois2010} and low-mass brown-dwarfs \citep[e.g.][]{chauvin2005,thalmann2009} that have been enabled by the advent of high-quality adaptive optics correction and sophisticated PSF subtraction techniques. This has allowed for various kinds of characterization of such systems \citep[e.g.][]{janson2010,bowler2010,currie2011,janson2011}, and opened up new domains for the types of planets that can be studied; for instance, observations within the disk gap of the transitional disk LkCa15 \citep[e.g.][]{espaillat2007,thalmann2010} using sparse aperture masking has recently revealed what could potentially be a planet in the process of forming \citep{kraus2011}.

Among these discoveries, one claimed planet detection that stands out as peculiar in many ways is that of Fomalhaut~b \citep[hereafter K08]{kalas2008}. The presence of a planet around Fomalhaut has been predicted on the basis of the geometry of the debris disk in the system \citep{kalas2005}, which has a sharp inner edge and a center that is offset from that of the star. This implies that it must be eccentric, which could in turn be indicative of the presence of a planetary companion (or several) exerting a gravitational influence on the disk \citep[e.g.][]{quillen2006,chiang2009}, although alternative interpretations do exist \citep{jalali2011}. Hence, when a point-source was discovered in two epochs within the disk gap (K08), with a direction of motion largely parallel to the disk edge, it was assumed that this was an image of the predicted disk-perturbing planet. However, the observed properties of the point-source are hard to consolidate with such an interpretation. Unlike the other detections mentioned above, which were made at near-infrared wavelengths where young substellar objects radiate the bulk of their energy, the Fomalhaut~b candidate is detected only at visible wavelengths, where the expected emission is near zero. No corresponding near-infrared radiation has so far been detected, despite several attempts \citep{kalas2008,marengo2009}. Several alternative interpretations of the observed properties have been made, which will be discussed in Sect. \ref{sec_discuss}. Regardless of interpretation, it remains the case that the best way to increase our understanding of the system and test whether a planet is associated with the observed point-source is to better constrain its near-infrared properties. 

Motivated by this, we have performed a study with the \textit{Spitzer Space Telescope} in order to improve the detection limits at 4.5~$\mu$m, which is the wavelength range where Fomalhaut~b is expected to emit its peak flux. We describe our observational methods and data reduction in Sect. \ref{sec_obs}, our various data analysis approaches and results in Sect. \ref{sec_result}, and discuss the implications for the Fomalhaut system in Sect. \ref{sec_discuss}.

\section{Observations and Data Reduction}
\label{sec_obs}

Our observations were taken with the Infrared Array Camera \citep[IRAC;][]{fazio2004} of the \textit{Spitzer Space Telescope} as part of program 70009, and consist of eight individual runs spread over cycle 7, from August 2010 through July 2011 (see Table \ref{t:obslog}). Each run consists of 48 exposures, structured as a cycle of a 12-point Reuleaux dither pattern with four exposures per position, which enables efficient spatial oversampling and bad pixel removal. The individual exposures have integration times of 10.4~s each with an execution time of 12~s, leaving the primary saturated in individual frames.  All observations were taken in the 4.5~$\mu$m band where the peak flux of Fomalhaut~b is expected. The observing strategy was optimized for angular differential imaging (ADI) purposes, with a large spread in telescope roll angles. Since active rotation around the optical axis of the telescope is not possible, we exploited the fact that nominal rotation occurs naturally over the course of the year and distributed the observations as uniformly over the observing cycle as the scheduling allowed. Given the observing windows for Fomalhaut, this led to observations being acquired in August and December of 2010, and January and July of 2011, with a range of position angles as summarized in Table \ref{t:obslog}.

\begin{table*}[p]
\caption{Log of Fomalhaut observations in \textit{Spitzer} program 70009.}
\label{t:obslog}

\begin{tabular}{lccccc}
\hline
\hline
AOR ID & Chan. & Exp. & Frames & MJD & PA \\
\hline
40250112 & 4.5~$\mu$m & 10.4~s & 48 & 55416.265 & 254.5$^{\rm o}$ \\
40249856 & 4.5~$\mu$m & 10.4~s & 48 & 55423.457 & 257.6$^{\rm o}$ \\
40249600 & 4.5~$\mu$m & 10.4~s & 48 & 55547.666 & 52.6$^{\rm o}$ \\
40249344 & 4.5~$\mu$m & 10.4~s & 48 & 55561.778 & 58.4$^{\rm o}$ \\
40249088 & 4.5~$\mu$m & 10.4~s & 48 & 55571.925 & 62.2$^{\rm o}$ \\
40248832 & 4.5~$\mu$m & 10.4~s & 48 & 55579.832 & 65.0$^{\rm o}$ \\
40248576 & 4.5~$\mu$m & 10.4~s & 48 & 55762.480 & 244.5$^{\rm o}$ \\
40250368 & 4.5~$\mu$m & 10.4~s & 48 & 55765.605 & 245.7$^{\rm o}$ \\
\hline
\end{tabular}
\end{table*}

Basic data reduction for all the observations was performed with the \textit{Spitzer} Science Center (SSC) IRAC Pipeline (version S18.18.0), which produced Basic Calibrated Data (BCD) frames and data quality masks for each individual exposure. We also used the post-BCD IRACproc package \citep[version 4.3;][]{schuster2006} for the sole purpose of removing outliers (cosmic-rays) for each frame. Subsequent steps were performed with custom procedures in IDL. An extra bad pixel removal step was introduced in order to identify and remove residual bad pixels that occurred only in single frames. This was done by identifying outliers from the median of each quadruplet of frames that were taken contiguously for a given dither position. The absolute center of the PSF in each frame was determined by cross-correlating the spider pattern with itself after a rotation by 180 degrees. All frames were then shifted to a common center and oversampled to a pixel scale of 300 mas/pixel, after which the ADI-based PSF subtraction could commence. This procedure is based on the Locally Optimized Combination of Images (LOCI) procedure \citep{lafreniere2007,lafreniere2009}, with adaptations to suit the new type of observational scheme and the science goals. The main aspects of the data set that distinguish it from most types of situations in which LOCI is applied are the extremely high PSF stability for space-based observations, the very large number of frames ($48 \times 8 = 384$), and the relatively small separation of the region of primary interest (compared to the PSF size, FWHM of 1.72\arcsec). Particularly the latter two factors make the data prone to substantial self-subtraction in a regular LOCI context. Since one of our main objectives in this study is to set a firm upper flux limit in case no detection would be made, we adapted the procedure in such a way as to avoid self-subtraction to a very high degree, while still maintaining a strong contrast performance. This is a conservative approach, and we note that it is entirely possible that the contrast could be even further enhanced with a more aggressive LOCI implementation. This will be a subject of future studies. 

Our adapted LOCI implementation follows the following procedure: First, each frame is sequentially subjected to an individual optimized reference PSF construction and subtraction. The basic optimization area is an annulus with an inner radius of 15 pixels and an outer radius of 60 pixels, centered on the star. This optimization area is split up into pieces of approximately 9-by-9 pixels ($\sim$1.5~FWHM). In practice, this is done in such a way that the basic annulus is split up into four annuli, each 9 pixels in width, and each annulus is split up azimuthally in such a way that an integer number of segments of equal angular width are created, and such that the length of the inner edge of the segment is as close to 9 pixels as possible. We will refer to these segments as `exclusion areas' henceforth. For each area, an individual LOCI optimization is executed where the optimization area consists of the full 15--60 pixel annulus, but with the exclusion area removed. Based on the resulting LOCI coefficients, a subtraction is then made only in the exclusion area. The resulting small area is saved to a frame which is put together piece by piece from the subtractions corresponding to the respective exclusion areas. Hence, in this procedure, the subtraction area is equal to the exclusion area, and is completely non-overlapping with the optimization area. The advantage of this approach is that it becomes impossible for the algorithm to systematically fit for any companion in the data, and thus the self-subtraction will be approximately zero. The possible cost comes from the fact that we exclude the stellar PSF regions that are physically the closest to the companion, and which may therefore correlate best with the actual PSF noise at the exact position of the companion. As we will see in the testing described in Sect. \ref{sec_result}, the procedure indeed works extremely well for avoiding any self-subtraction.

For the procedure described above, the reference frames are chosen from the available library of frames based on how far separated the position angles are between the subtraction frame and a given reference frame. Since we wish to ensure a very low degree of self-subtraction of any real companion, we conservatively choose that the separation must be at least 1.72\arcsec (1~FWHM) at the inner edge of the exclusion area. This corresponds to different position angles at different separations from the star, hence the subtraction areas at larger separation typically have access to a larger number of reference frames. However, since the position angles are spread over a large range (see Table \ref{t:obslog}), every examined position has access to a sufficient number of reference frames for a very high quality PSF subtraction.

Once all frames have been subjected to the PSF subtraction, they are de-rotated to a common sky orientation where North is up and East is to the left, and collapsed into a final frame using the median of the individual frames. Another step is then performed, in order to take boundary effects into account. If a real companion happens to be positioned exactly on the boundary between two or more exclusion zones, it is partially vulnerable to residual oversubtraction even in the above procedure. To overcome this effect, we perform three additional LOCI procedures in exactly the same way as described above, with only one difference: In the first additional procedure, the exclusion zones are shifted one half step in the azimuthal direction, in the second one, they are shifted one half step in the radial direction, and in the final one, they are shifted half a step in each direction. In this way, four different reduced frames are available to check if some companion gets fainter due to boundary effects in some of the frames, and additionally allows to check for spurious features that could occur in some reductions but not in others. In this case there is a box-like artefact in two of the images but otherwise they all show nicely consistent patterns, hence for the further analysis we use the mean of the frames from the four reductions. 

In order to evaluate the achieved contrast as a function of separation, we use the same procedure as described in \citet{marengo2009}, evaluated from the standard deviation in consecutive 1-pixel annuli. An important difference is that we always use a 5$\sigma$ criterion here for our measurements, rather than 3$\sigma$ as in \citet{marengo2009}. This is more stringent in the presence of speckle noise, although note also that there is a fairly close equivalence between a 5$\sigma$ single-signature criterion and a 3$\sigma$ double-signature criterion \citep{janson2008}, the latter of which is relevant for the \citet{marengo2009} data.

\section{Results and Analysis}
\label{sec_result}

\begin{figure*}[p]
\centering
\includegraphics[width=16cm]{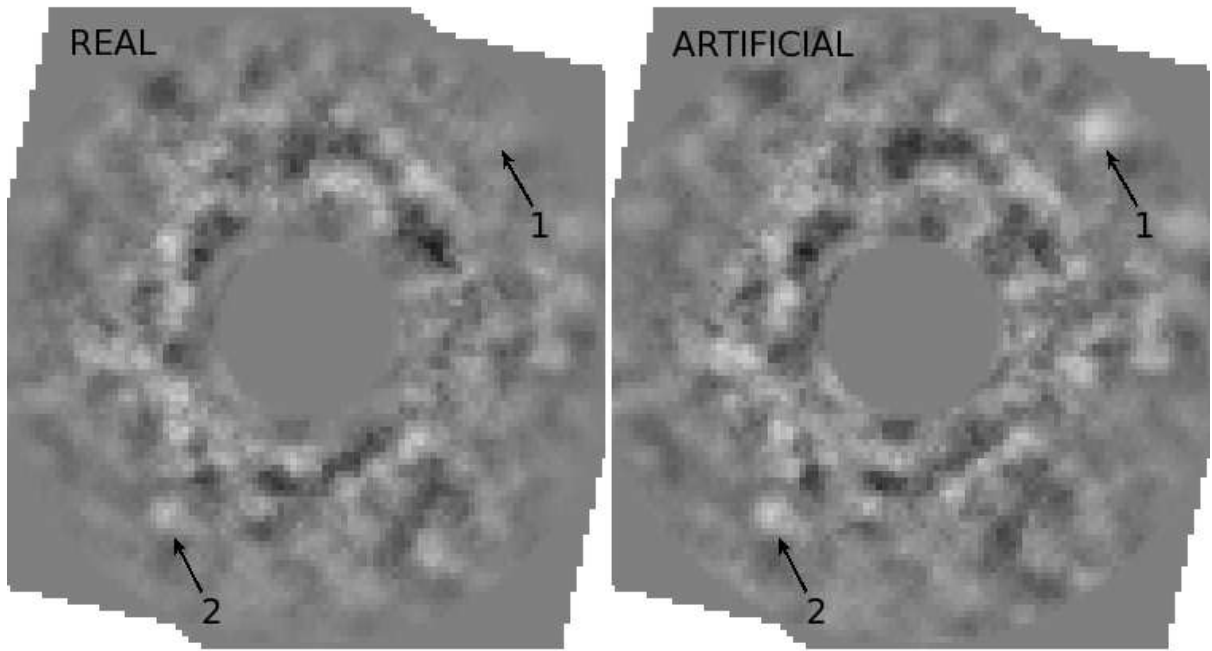}
\caption{Final reduced image for the real data (left) and for the data with an artificial companion introduced at the expected position of Fomalhaut~b (right). Arrow 1 points out the expected position of the companion based on earlier detections in the visible-light images. There is no corresponding point source seen in the real data. The artificial companion in the right-hand image passes through the data reduction without any flux loss, verifying that the non-detection is real, such that a stringent upper flux limit can be set. Arrow 2 points toward the brightest possible point source in the field. Its position is consistent with a ring-nested orbit, but the significance is too low to make any assessment of whether or not it is a real object. North is up and East is to the left in the images.}
\label{f:fomaltile}
\end{figure*}

\begin{figure*}[p]
\centering
\includegraphics[width=12cm]{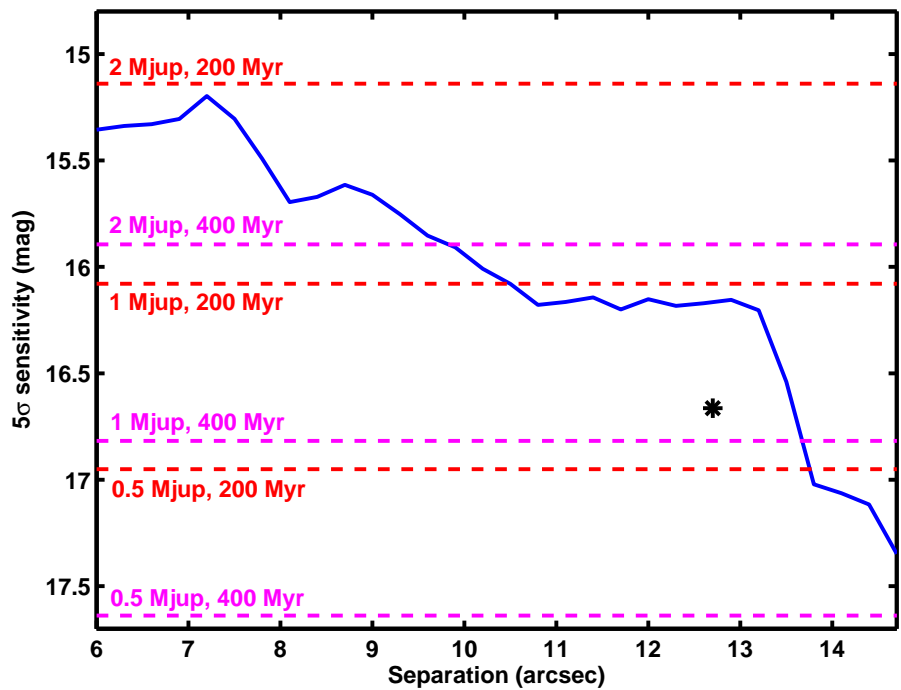}
\caption{Sensitivity limit as function of separation from Fomalhaut. The solid line is the azimuthally averaged sensitivity profile. There is a bump around 12\arcsec due to concentrated noise at some ranges of position angles. The expected position of Fomalhaut~b from the visible wavelength range detections is in a cleaner part of the image space, and the local sensitivity at this position is shown by the black asterisk. Also plotted as dashed lines are the expected brightnesses for planets with a few different combinations of mass and age, according to models from \citet{spiegel2011}.}
\label{f:contrast}
\end{figure*}

We show the final reduced image in Fig. \ref{f:fomaltile} and the corresponding sensitivity limits in Fig. \ref{f:contrast}, where we also plot the expected brightnesses for planets of a few different masses at ages of 200 and 400~Myr. A very substantial improvement in contrast performance is achieved with our LOCI implementation, with an order of magnitude improvement in flux detectability compared to the conventional ADI reduction in \citet{marengo2009}, and with a more stringent detection criterion. No signature is found at the position where Fomalhaut~b would be expected. Hence, we estimate an upper limit (5$\sigma$) at the relevant position based on the standard deviation in a 7-by-7 pixel box ($\sim$1.2 FWHM on the side). This gives an upper limit of 16.7~mag, which corresponds to 38.8~$\mu$Jy, again more than an order of magnitude improvement over previous data. 

In order to test that the non-detection is real and not an effect of any unexpected oversubtraction in the LOCI procedure, we make a full reduction following the exact same procedure as described in the previous section, except we also introduce a faint artificial companion (a Gaussian with 1.72\arcsec FWHM) in all the pre-LOCI frames, at the expected position of Fomalhaut b, with a flux of 57~$\mu$Jy (this corresponds to an effective temperature of $\sim$250~K, or equivalently 0.5--1~$M_{\rm jup}$ at 200~Myr and 1--2~$M_{\rm jup}$ at 400~Myr). The artificial companion passes through the LOCI reduction entirely unaffected, with the same measured flux in the final frame as the flux that was put in within the error bars, and is well visible at 7$\sigma$ confidence in the final frame (see Fig. \ref{f:fomaltile}). Note that the introduction of a new feature in the data affects the LOCI reduction itself -- since the artificial companion exists in most reference frames as an additional feature that has to be fit for by the algorithm, the fit quality will in general be slightly worse on average. Since the extent of the companion is small with respect to the optimization area and it is faint with respect to the stellar PSF, the effect is small but noticeable as a marginally higher general noise level in the final reduced frame. In summary, the procedure described here validates that the non-detection is real, and thus the upper flux limit is relevant. 

\begin{figure*}[p]
\centering
\includegraphics[width=16cm]{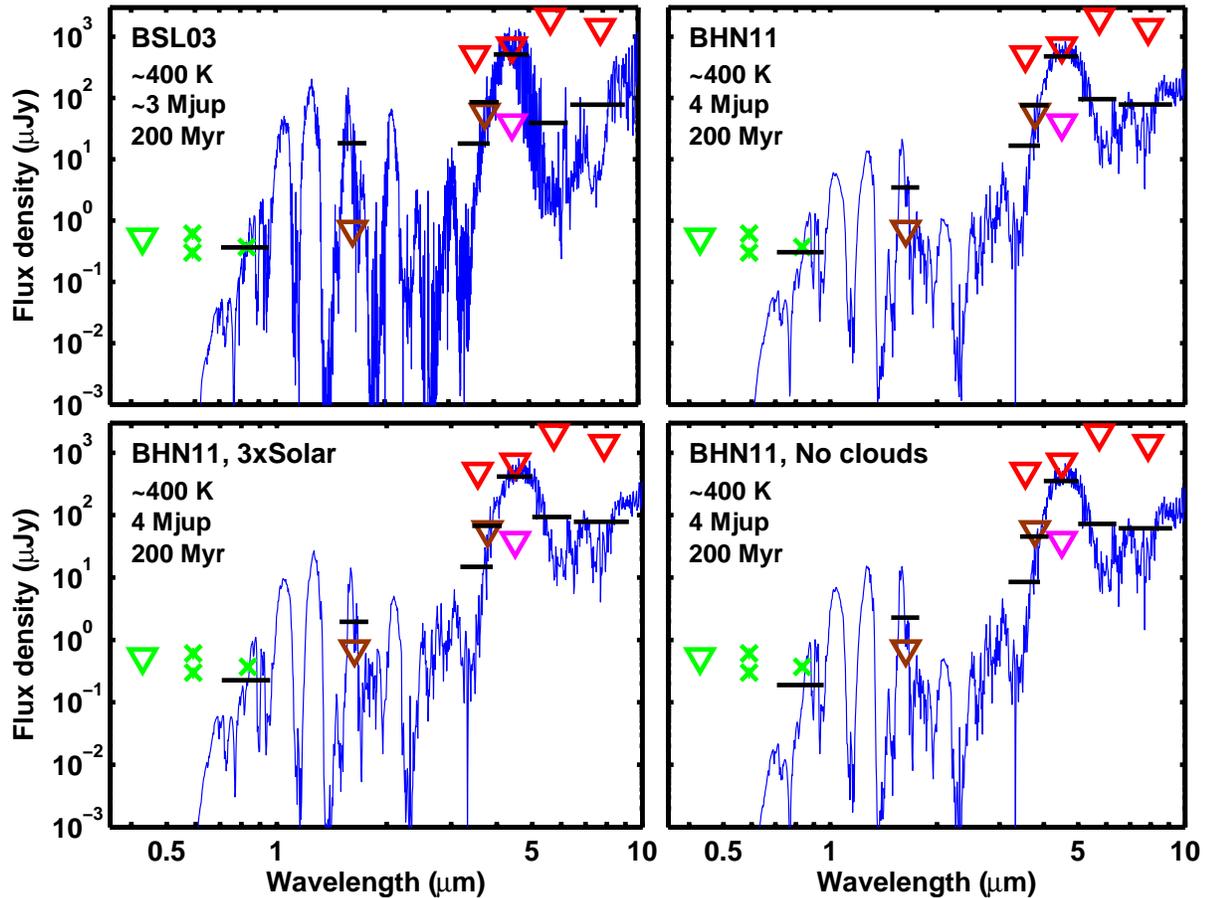}
\caption{Model comparisons to the observational data for a $\sim$400~K atmosphere ($\sim$2--3~$M_{\rm jup}$ at $\sim$200~Myr for BSL03, 4~$M_{\rm jup}$ at 200~Myr for BHN11). The solid line in each panel is the model spectrum, and the black lines are the corresponding fluxes in the respective photometric bands. Crosses mark detections and triangles mark upper limits. HST data points from K08 are in green and Keck/Gemini data points from K08 are in brown. Note that the error bars at F606W and F814W are smaller than the sizes of the symbols. Red symbols are \textit{Spitzer} upper limits from \citet{marengo2009}. Our new upper limit from \textit{Spitzer} is the magenta triangle. Upper left: A BSL03 model. Upper right: A BHN11 model with patchy clouds and Solar composition. Lower left: A BHN11 model with patchy clouds and increased metallicity. Lower right: A BHN11 model with Solar composition and a cloud-free atmosphere. Models in this effective temperature range are required to produce an adequate amount of flux in F814W, but they are typically inconsistent with the upper flux limits in H-band and/or L'-band, and always fully inconsistent with the upper limit at 4.5~$\mu$m}.
\label{f:400k}
\end{figure*}

We show our upper limit in Fig. \ref{f:400k}, along with other upper limits from the literature at various wavelengths, as well as the detection values in the visual wavelength range. We also compare these values with various theoretical models. One model spectrum is from \citet{burrows2003} (henceforth BSL03), corresponding to a 2--3~$M_{\rm jup}$ planet (interpolated between 2~$M_{\rm jup}$ and 5~$M_{\rm jup}$ to match the flux value at F814W\footnote{F606W and F814W are filters for the Advanced Camera for Surveys on the \textit{Hubble Space Telescope}, centered on wavelengths of $\sim$600 and $\sim$800~nm, respectively}) at $\sim$200~Myr. As noted in K08 and \citet{marengo2009}, this model is similar to e.g. the \citet{fortney2008} model except in H-band where the BSL03 flux is higher. It can be seen from the comparison to the data that in the context of this model, a thermal flux interpretation is entirely inconsistent with both the non-detection in H-band and our non-detection in the IRAC 4.5~$\mu$m channel. The H-band flux is strongly model-dependent as it is sensitive to uncertainties in opacity and the treatment of clouds. However, by contrast, the 4.5~$\mu$m flux is very insensitive to these effects, and varies only very marginally across different models. To show this, we use newer models from \citet{spiegel2011} based on \citet{burrows2011} (henceforth BHN11), which has improved opacities and cloud treatment, and also includes a set of entirely cloud-free models for comparison. 

By using these new models and including clouds, it is possible to suppress the H-band flux to a significant extent (though typically still not quite to a sufficient extent for a non-detection to be consistent with thermal flux in F814W). However, this is not the case at 4.5~$\mu$m. The flux remains very stable for a constant effective temperature (typically $\sim$400~K for these model comparisons), regardless of how the clouds are treated including the cloud-free case, independently of opacity treatment and also of specific metallicity (the BHN11 models provide both Solar and super-Solar metallicity cases). We conclude that the effective temperatures required to get any substantial contribution of thermal flux to the observed F814W data point are simply inconsistent with the non-detection at 4.5~$\mu$m.

\begin{figure*}[p]
\centering
\includegraphics[width=16cm]{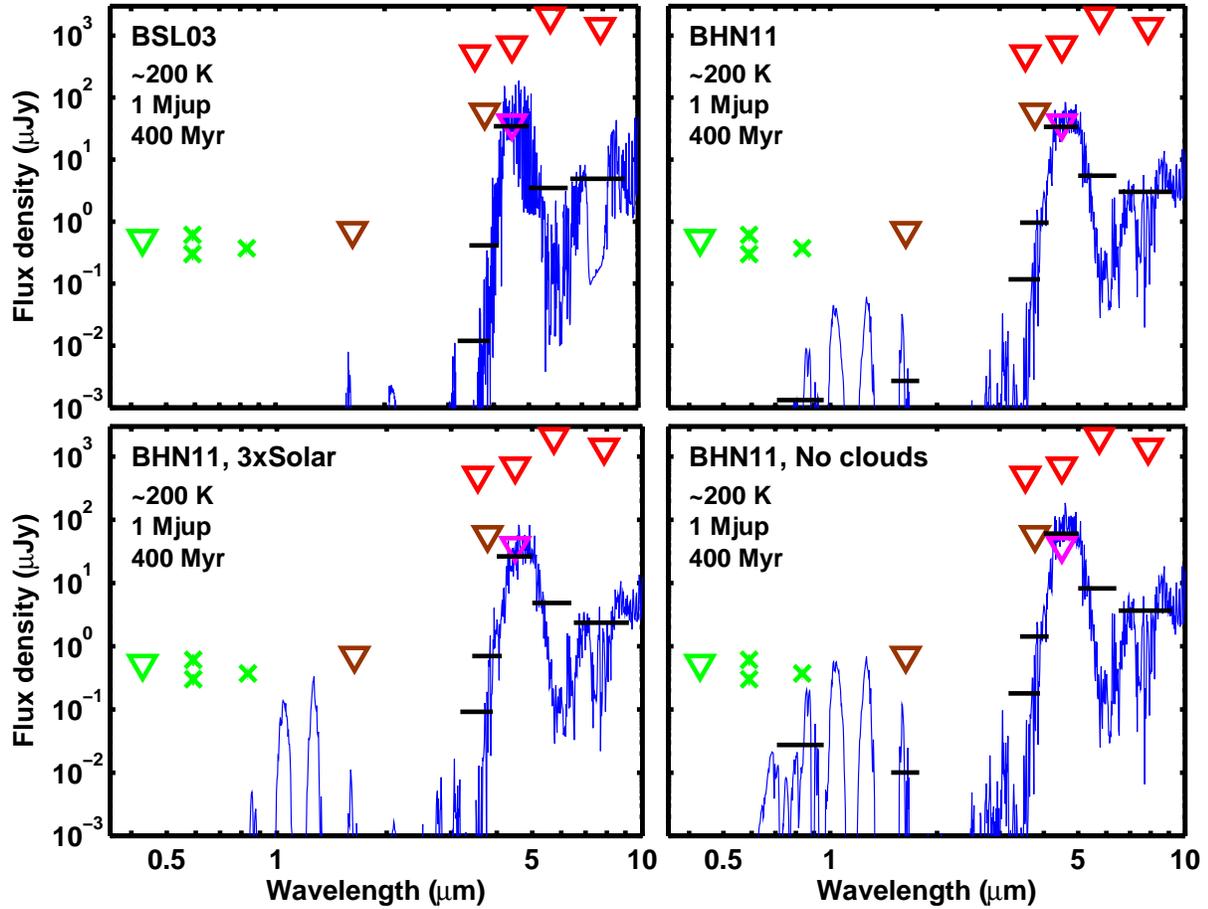}
\caption{Model comparisons to the observational data for a $\sim$200~K atmosphere (1~$M_{\rm jup}$, 400~Myr). The symbols have the same meaning as in Fig. \ref{f:400k}. Upper left: A BSL03 model. Upper right: A BHN11 model with patchy clouds and Solar composition. Lower left: A BHN11 model with patchy clouds and increased metallicity. Lower right: A BHN11 model with Solar composition and a cloud-free atmosphere. These models are marginally consistent with the upper flux limit at 4.5$\mu$m, but predict one to several orders of magnitude too little flux to be consistent with thermal radiation in F814W.}
\label{f:200k}
\end{figure*}

In order to comply with the upper limit at 4.5~$\mu$m, we need effective temperatures in the range of $\sim$200~K or lower, corresponding to, e.g., 1~$M_{\rm jup}$ at 400~Myr (see Fig. \ref{f:200k}). As a side point, this latter age would correspond to a new but so far unpublished estimate (E. Mamajek, priv. comm.) which is a bit older than the mean estimate of 200~Myr used in K08 and \citet{marengo2009}. Here we do not make any assessment of the relative credibility of these two estimates, but simply remark that it has no real relevance for the spectral comparisons we are performing here. The main factor (beyond cloud treatment and opacity) that affects the spectral energy distribution is the effective temperature. The mass to which this temperature corresponds depends on the age and vice versa, but this has little impact on the spectrum, especially for the small discrepancy in age that we are concerned with here. Hence, as long as we do not actually try to determine the mass, we do not need to assess which of 200~Myr or 400~Myr is the better estimate. While on this note, it is worth pointing out that hot/warm/cold-start models are of no significant relevance to this discussion, since convergence will have occurred at these ages and masses, regardless of initial entropy \citep{spiegel2011}.

For the colder models required to match the 4.5~$\mu$m upper flux limit, virtually all flux at shorter wavelengths is lost, and there is no way to match the F814W point with any thermal flux. The closest case is the extreme case of a completely cloud-free atmosphere and solar metallicity, but also in this case the flux is more than an order of magnitude too small (see Fig. \ref{f:200k}). Increasing metallicity has the effect of decreasing flux at 800~nm compared to 4.5~$\mu$m, hence decreasing metallicity could have the opposite effect. However, Fomalhaut~A has a super-solar metallicity with a mean measured value of 0.3~dex in a compilation of literature values in \citet{soubiran2010}. Fomalhaut~b should have an equal or larger metallicity -- thus, adjusting the metallicity is not a feasible route toward reaching consistency with the observational data.

Based on the above results and considering the further aspects of the collected body of observations of Fomalhaut as will be discussed in detail in Sect. \ref{sec_discuss}, it is highly unlikely that the observed flux at visible wavelengths has any direct connection to the suspected giant planet that might shepherd the debris disk of Fomalhaut and force it into an eccentric state \citep{kalas2005}. In this context, and considering our very strong detection limits, it is interesting to assess whether this shepherding planet (what might be referred to as the `real' Fomalhaut b) can be seen in our images. Since the shepherding planet can be as low in mass as 0.5~$M_{\rm jup}$ \citep{chiang2009}, it is fully plausible that it could remain undetected in our images if the age is as old as 400~Myr, and since its orbit covers a range of projected separations, it can also hide in some parts of the orbit even if the mass is slightly higher. However, we do cover a very large fraction of its possible parameter space, and it is noteworthy that the brightest possible point-source in the field with a significance of 4.3$\sigma$ is in fact located at a position that would be consistent with a ring-nested orbit (arrow 2 in Fig. \ref{f:fomaltile}). However, the fact that the possible point source is not at a 5$\sigma$ confidence level obviously means that more data or an even further improved PSF reduction would be necessary in order to test its validity.

\section{Discussion}
\label{sec_discuss}

In this section, we discuss in detail how our non-detection at 4.5~$\mu$m affects the interpretation of Fomalhaut~b, and what can be deduced about the system from the full body of existing data. The data points that exist for the detected point-source in K08 are two detections in the F606W filter from 2004 and 2006, and one detection in the F814W filter from 2006. These data points are shown in Figs. \ref{f:400k} and \ref{f:200k}. Note that the object is variable between 2004 and 2006. This is not a small effect; in fact, it corresponds to a change in brightness (dimming) of a factor 2, at a confidence of 8$\sigma$. This is the same level of confidence as, e.g., the second-epoch F606W detection of the point-source altogether. Hence, to the extent that we can trust the data at all, we must consider this variability as a real effect, and it needs to be accounted for in a comprehensive interpretation of the object. In addition to these data, a third epoch HST observation has also been acquired but has not yet been published at the time of writing (P. Kalas, priv. comm.). There are also upper flux limits from non-detections in a range of bands in \citet{kalas2008}: F435W, H, CH4S, CH4L, and L', and upper limits at 3.6, 4.5, 5.8 and 8.0~$\mu$m from \textit{Spitzer} \citep{marengo2009}, the second of which we improve on in this article.

There are two main lines of interpretation of the point source in K08, only one of which actually includes any flux from a planet. In this scenario, the flux at F814W originates from the thermal emission of a planet, which has to be close to a mass of $\sim$3~$M_{\rm jup}$ in order to fit the data point assuming an age of 200 Myr \citep[e.g.][]{burrows2003,fortney2008}. This is however poorly consistent with the rest of the available data. Most obviously, it is a factor 20--40 brighter in F606W than expected in such a scenario. In order to explain the F606W data in terms of brightness and variability, K08 infer a hypothesis of H$\alpha$ accretion on the planet. Given that this would require gas accretion at the same rate of a few Myr old TTauri stars like GQ Lup (accretion rate calculated in the supplemental material of K08), whereas Fomalhaut is 200--400~Myr and has no other known signs of gas anywhere in the system, we consider this hypothesis highly unlikely. 

Aside from this, there is the issue that a $\sim$3~$M_{\rm jup}$ companion should also be detectable in the near-infrared. This was an issue already in K08, where the H-band flux predicted by theoretical models was significantly higher than the upper limit from observations, and became an even larger issue given the \textit{Spitzer} data published by \citet{marengo2009}, where the additional limits at larger wavelengths provided very little opportunity for any flux to remain undetected from such a companion. As was shown in the previous section, our new \textit{Spitzer} data now provides even much tighter constraints on the thermal emission hypothesis, and provides the opportunity to conclusively address this issue. We find that any thermally emitting companion responsible for the F814W flux (effective temperatures of $\sim$400~K, e.g. 2--3~$M_{\rm jup}$ at 200~Myr) would have emitted more than an order of magnitude more flux at 4.5~$\mu$m than our 5$\sigma$ upper limit, regardless of the choice of theoretical models. Conversely, any companion that thermally emits radiation at 4.5~$\mu$m at levels comparable to our upper limit (effective temperatures of $\sim$200~K, e.g. 1~$M_{\rm jup}$ at 400~Myr) would emit at least an order of magnitude too little flux at F814W to explain the observations, and for most realistic model parameters (e.g. any inclusion of clouds) the flux would be even much smaller. Hence, we can firmly exclude the hypothesis that any of the observed flux in K08 actually originates from a giant planet.

Given that the SED of the K08 point source can be interpreted as having a reasonable match to the stellar SED, it seems more likely that what is seen in the K08 images is some form of reflected or scattered radiation from the star. Given the large effective area that is required for this, the only plausible origin for reflected/scattered radiation is dust. Hence, it is likely that we are seeing a concentration of dust, which may or may not be associated with a planet. We will discuss some dust-related interpretations in the following: firstly, we will consider the hypothesis of an optically thick disk around a giant planet. This is the second preferred scenario in K08, and requires a disk radius of at least $\sim$20~$R_{\rm jup}$ (for the case of high albedo; larger for the case of low albedo). It may not be unreasonable that circumplanetary rings start out with such properties. However, it might also be argued that the age of the Fomalhaut system of 200--400~Myr should have provided adequate time for moons to form and excavate the disk, leaving only rings of a much smaller effective size. Regardless of whether the optically thick disk scenario is fundamentally realistic or not, there are several reasons for why this scenario is inconsistent with the existing data. One very important constraint is that such a disk cannot account for the factor 2 variability observed in F606W. Furthermore, the spin of the star has been recently measured with interferometry \citep{lebouquin2009}. If the spin of the star is aligned with the plane of the disk, this means that the fainter Western side is closer to us than the brighter Eastern side. Although this is opposite to what would be expected for purely forward-scattering dust, it is shown in \citet{min2010} to be consistent with the scattering behaviour of large dust grains. If it is indeed the case that the Western side of the disk is closer to us, and the K08 point source orbits within the disk, then it follows that the object is located between the parent star and Earth, in the radial direction. It would be very difficult for an optically thick disk to reflect large amounts of light to Earth under such circumstances. 

In addition to these points, another important argument against the involvement of a giant planet (which also applies to the thermal emission case discussed above) is the orbit of the object. The third epoch astrometric measurement implies a ring-crossing orbit for the point source (P. Kalas, priv. comm.). Although this is not yet published, it would not be surprising given the previously published data, because, in fact, already the first two epochs are inconsistent with an orbit that traces the edge of the ring, at a $\sim$2$\sigma$ level (the direction is largely consistent with such an orbit, but the speed is not). This is pointed out by \citet{chiang2009}, who do not put a large emphasis on this fact, as they argue that the error bars are probably underestimated. However, we note that this is diametrically opposite to the interpretation of the astrometric errors in K08, where it is concluded that they are an upper limit to the real error (K08, supplemental material). A ring-crossing orbit would be inconsistent with an association of the point source to a giant planet, as it would strongly affect the geometry of the disk, hence any planet associated with the point source would need to be low in mass. 

In the context of a low-mass planet, we note that one way to suppress the 4.5~$\mu$m flux with respect to shorter wavelengths is to consider hotter temperatures and smaller surfaces. Effective temperatures higher than the $\sim$200--400~K that we have considered thus far are not reasonable for isolated objects at the system age, particularly for smaller (and thus lower-mass) objects than Jupiter-class planets. However, following intense bombardment of planetesimals, rocky protoplanets may acquire molten surfaces and reach effective temperatures of $\sim$1000--3000~K over brief periods of time \citep{kempton2009}. Hence, the possibility that the observed light-source could be a hot collisional afterglow of a $\leq$10~$M_{\rm Earth}$ object should not be dismissed out of hand. Still, it would probably be difficult to reproduce all the observed data points in such an interpretation, particularly the detection in F606W, and the simultaneous non-detections in H-band and at 4.5~$\mu$m. The \citet{kempton2009} models do not cover the full relevant wavelength range, but if we consider, for example, a planet of 10~$M_{\rm Earth}$ and 1.8~$R_{\rm Earth}$ and work from pure blackbody considerations, we can establish that the brightness temperature required to reproduce even the lowest of the F606W data points (0.30~$\mu$Jy) is more than 1500~K, whereas the upper limits in H-band (0.71~$\mu$Jy) and at 4.5~$\mu$m (38.8~$\mu$Jy) both require brightness temperatures of $\sim$700~K or less. We consider that future modelling efforts would be worthwhile to examine whether these conditions and the rest of the flux limits can all be simultaneously fulfilled, but for the purpose of our discussion here, we simply treat it as an option that cannot be categorically excluded. On balance, it should be noted that an observation of this type of scenario is probably rather unlikely, for several reasons, including the fact that it is expected to last over timescales of $\sim$10$^4$~yr for the case of a thin atmosphere, very short compared to the age of Fomalhaut. The timescale can be extended if the atmosphere is thickened and clouds are included, but the observable brightness temperature decreases accordingly. As a side note, if clouds were involved, they could possibly account for the variability in F606W in this scenario.

Given the SED of the point source and its variability, along with the considerations above, the perhaps most plausible way to consistently explain the observed properties of the K08 point source is through a cloud of dust, which is either transient or has a transient component. There are two possible scenarios associated with such an interpretation that have been suggested in the literature. In one scenario, the observed point source is a residual (gradually dispersing) dust cloud from a recent planetesimal collision. We certainly know that such collisions should occur frequently in the Fomalhaut system, given that they are the very origin of dust in debris disks. This scenario is mentioned by K08, who argue against it based on the fact that such collisions should be much more common within the actual ring feature than just outside of it where the point source is observed, hence the relative probability to observe it where it is observed should be low. This is certainly true, but we note that there is a clear selection effect involved -- due to the high visual brightness of the ring and the speckle-like nature of the noise, any number of equivalent events that hypothetically do happen within the ring feature would be likely to pass unnoticed. One might also hypothetically imagine that the cloud is in the present position for some specific dynamical reason, for instance if the material is trapped in resonance with a giant planet situated elsewhere.

The second scenario is essentially the same as the first, but involves a central rocky/icy object with a mass less than $\sim$10~$M_{\rm Earth}$, to which a swarm of planetesimals is gravitationally bound \citep{kennedy2011}. Collisions between these planetesimals produce the observed dust. We consider both of these scenarios to be reasonable within the constraints set by the data, and simply conclude that the K08 point source is well consistent with a transient or semi-transient dust cloud, which may or may not be gravitationally bound to a central object of planetary mass. With regards to the fact that the point source has been frequently referred to as a directly imaged planet in the literature, we note that this is incompatible with the observational evidence, for two independent reasons: (1) Although it cannot be formally excluded, there is insufficient evidence to support that there is any compact object of planetary mass associated with the point source altogether. (2) Even if such an object is present, in several of our considered scenarios we do not observe any photons from this object itself, hence it cannot be established that it has been directly imaged. 

\section{Conclusions}

In this paper, we have presented observations performed with \textit{Spitzer}/IRAC in the 4.5~$\mu$m band for the purpose of trying to detect thermal emission from Fomalhaut~b. A new LOCI-based PSF subtraction scheme was implemented to achieve high contrast with minimal companion flux loss, which enabled an order of magnitude improvement in contrast-limited sensitivity with respect to previous efforts. The non-detection of any flux at the expected position can therefore be used to provide strong constraints on the underlying physics of the point-source seen at visible wavelengths. In particular, we find that there is almost certainly no direct flux from a planet contributing to the visible-light signature. This, in combination with the existing body of data for the Fomalhaut system, strongly implies that the dynamically inferred giant planet companion and the visible-light point source are physically unrelated. This in turn implies that the `real' Fomalhaut~b still hides in the system. Although we do find a tentative point source in our images that could in principle correspond to this object, its significance is too low to distinguish whether it is real or not at this point. 

Concerning the visible-light point source, its underlying physics is unclear, but the only hypothesis that can be shown to reasonably fit all existing data is an optically thin dust cloud, which is transient or has a transient component. If this interpretation is valid, the cloud may or may not be physically bound to a central object in the super-Earth mass regime.  

\acknowledgements
The Fomalhaut system is a rich topic of conversation, and we thank Adam Burrows, Carsten Dominik, James Graham, Ray Jayawardhana, Paul Kalas, Michiel Min, and many others for interesting discussions. This work is based on observations made with the \textit{Spitzer Space Telescope}, which is operated by the Jet Propulsion Laboratory, California Institute of Technology under a contract with NASA. M.J. is funded by the Hubble fellowship. J.C.C., J.R.B., and P.W. were supported by grant AST-1009203 from the National Science Foundation. D.S.S. gratefully acknowledges support from NSF grant AST-0807444 and the Keck Fellowship.

{\it Facilities:} \facility{Spitzer}.

\clearpage

\end{document}